\documentclass[usegraphicx,usenatbib]{mn2e}
\usepackage{amsmath}
\title[Theoretical Modeling of Emission-Line galaxies]{Theoretical Modeling of Emission-Line galaxies: New Classification Parameters for Mid-Infrared and Optical  Spectroscopy}
\author[M. Mel\'endez et al.]{M. Mel\'endez, $^{1}$\thanks{Present address: Astronomy Department, University of Maryland, College Park, MD 20742}\thanks{E-mail:
marcio@astro.umd.edu} T. M. Heckman, $^{1}$ M. Mart\'inez-Paredes, $^{2}$\newauthor 
 S. B. Kraemer $^{3}$ and C. Mendoza $^{4}$\\
$^{1}$Department of Physics and Astronomy, Johns Hopkins University, Baltimore, MD, 21218, USA\\
$^{2}$Instituto Nacional de Astrof\'isica, \'Optica y Electr\'onica, Puebla, 72840, Mexico\\
$^{3}$Department of Physics, The Catholic University of America,Washington, DC 20064, USA\\
$^{4}$Centro de F\'isica, IVIC, PO Box 20632, Caracas 1020A, Venezuela }

\begin{document}

\maketitle

\begin{abstract}
  We have carried out extensive and detailed photoionization modeling to successfully constrain the locations of different emission-line galaxies in optical and mid-infrared diagnostic diagrams. Our model grids cover a wide range in parameter space for the active galaxy continuum and starburst galaxies with different stellar population laws and metallicities. We compare the predicted AGN and star-formation mid-infrared line ratios  [Ne~III]15.56$\umu$m/[Ne~II]12.81$\umu$m and [O~IV]25.89$\umu$m/[Ne~III]15.56$\umu$m  to the observed values, and find that the best fit for the AGN is via a two-zone approximation. This two-zone approximation is a combination of a matter-bounded component, where [Ne~III] and [O~IV] are emitted efficiently, and a radiation-bounded component that maximizes [Ne~II] emission. We overlay the predictions from this two-zone approximation onto the optical [O~III]$\lambda$5007/H$\beta$ and [N~II]$\lambda$6583/H$\alpha$ diagnostic diagram derived from the Sloan Digital Sky Survey, to find that the high-density and low-ionization radiation-bounded component in our two-zone AGN approximation model provides a good lower limit for [N~II] emission. This establishes a new theoretical demarcation line for the minimum AGN contribution in this diagram. This new classification results by a factor of $\sim$ 1.4 in a higher AGN population than predictions derived from previous divisions of star-forming galaxies. Similarly, we define a maximum AGN contribution in the [O~III]/H$\beta$ and [N~II]/H$\alpha$ diagram by using a two-zone approximation within a parameter range typical of the narrow-line region.

\end{abstract}


\section{Introduction}

With the advent of extensive and accurate astronomical surveys in the last few years, there has been a rapid increase of astronomical data that facilitates large-scale statistical studies. For example, the Sloan Digital Sky Survey (SDSS) contains  more than 2 million optical spectra of galaxies and quasars and a few hundred thousand  optical and infrared stellar spectra \citep{2014ApJS..211...17A}. Given this extraordinary amount of observations, it is clear that efficient procedures must be devised to separate and classify the different objects that populate these surveys.  An approach is to classify the sources in terms of the principal excitation mechanisms of the line emitting gas. Historically, the most common line discriminant in the optical among the excitation mechanisms is the combination of the   [O~III]$\lambda$5007/H$\beta$ and [N~II]$\lambda$6583/H$\alpha$ ratios \citep[e.g.,][]{1981PASP...93....5B,1987ApJS...63..295V}. These ratios can separate photoionization by active star formation (hot stars) from that by the ``non-thermal"  power-law continuum of an active galactic nucleus (AGN). These demarcation lines can be defined empirically \citep[e.g., ][hereafter Ka03]{2003MNRAS.346.1055K} or via photoionization modeling of the emission-line spectra \citep[e.g., ][hereafter Ke01]{2001ApJ...556..121K}.  A further separation can be performed between Seyfert galaxies and low-ionization emission-line regions  \citep[][]{2006MNRAS.372..961K}. However, despite their great success,   the identification of the most heavily dust-obscured systems remains a challenge for optical diagrams \cite[e.g.,][]{2007ApJ...667..149F,2008ApJ...677..926S,2009MNRAS.398.1165G,2014arXiv1405.7940C}.

Similar diagnostics in the mid-infrared, which rely on emission from the narrow line region (NLR, Seyfert 1 galaxies show both broad- and narrow-line emission whereas Seyfert 2 galaxies only show narrow-line emission due to orientation effects), have proven to be excellent tools to separate the different galaxy excitation mechanisms. Mid-infrared diagnostics are more suitable to study dust-enshrouded systems where the effect of dust obscuration can hamper the interpretation of traditional optical diagnostics \citep[e.g.,][]{2008ApJ...682...94M,2008ApJ...678..686A,2009ApJ...700.1878R,2009MNRAS.398.1165G,2014arXiv1405.0670D}. In general,   it is important to understand  the underlying excitation mechanism in dust obscure systems  because they represent  a common phase in the evolution of most galaxies, especially in the early Universe at the peak of the dusty star formation and AGN activity  \citep[e.g.,][]{1972ApJ...178..623T,2007ApJ...667..149F,2009ApJS..182..628V,2014arXiv1405.7940C}.  Mid-infrared spectra provide access to important emission lines with a wide range of critical densities and ionization potentials that allow for a precise characterization of the physical properties and excitation mechanisms in the galaxy line-emitting regions. In this respect, low-ionization lines such as [Ne~II]12.81$\umu$m, which requires an ionization energy of E$_{IP}$ = 21.57~eV to be produced, are common in star forming galaxies \citep[e.g.,][]{2007ApJ...658..314H}. This line is well correlated with other star-formation indicators such as polycyclic aromatic hydrocarbon (PAH) features and the far-infrared continuum  \citep[e.g., ][]{2008ApJ...689...95M}. On the other hand, high-ionization lines, e.g. [O~IV]25.89$\umu$m (E$_{IP}$ = 54.94~eV) and [Ne~V]14.32$\umu$m/24.32$\umu$m (E$_{IP}$ = 97.12~eV), are  good indicators of the AGN power \citep[e.g.,][]{2008ApJ...682...94M,2009ApJ...700.1878R,2010ApJ...725.2270P,2010ApJ...716.1151W}. In addition, intermediate-ionization lines such as [Ne~III]15.56$\umu$m (E$_{IP}$ = 40.96~eV) can provide a unique perspective into composite systems where the AGN coexists with active star formation in the host galaxy \citep[e.g.,][]{2008ApJ...689...95M,2010ApJ...725.2270P,2012ApJ...758....1L}.

\citet{2010ApJ...716.1151W} (hereafter W10) presented a mid-infrared diagnostic that utilizes the most common emission lines observed in star-forming and Seyfert galaxies. The W10 [Ne~III]/[Ne~II] and [O~IV]/[Ne~III] diagnostic diagram shows a clear separation between AGN, star-formation, and blue compact dwarf (BCD) and Wolf-Rayet (WR) galaxies. This diagram can be quantitatively separated into three large sectors. The majority of the hard X-ray selected AGN from the {\it Swift} Burst Alert Telescope survey (BAT AGN) are located in the upper-right region where the [Ne~III]/[Ne~II] and [O~IV]/[Ne~III] ratios are greater than unity; for the rest of this paper we will refer to these as AGN dominated sources. Most of the star-forming and starburst galaxies are in the lower-left region with [Ne~III]/[Ne~II] and  [O~IV]/[Ne~III] ratios less than unity, while extreme starburst galaxies are located in the upper-left region with [Ne~III]/[Ne~II] $>1$ and  [O~IV]/[Ne~III] $<1$. Note that the selection process for the   BAT AGN depends only on the galaxy hard X-ray flux (14-195 keV) thus providing an unbiased sample of Compton-thin AGN. Moreover, because of its high-energy selection, the sample is also unbiased toward  star formation activity in  the host galaxy; therefore, sources in the BAT sample have a wide range of stellar contribution to their mid-infrared spectra. \citet{2009ApJ...704.1159H} (hereafter H09)  presented similar mid-infrared diagnostics  where, again,  the AGN and composite systems occupy a different branch to that from pure, extreme  starburst galaxies.

Despite the capabilities of these mid-infrared diagrams for distinguishing excitation mechanisms, there are few theoretical studies that attempt to define demarcation lines between activity types in galaxies. On the other hand, there are several studies that have derived theoretical classifications for star-forming galaxies and AGN based on optical diagnostic diagrams; however, these studies have been limited to the modeling of star-forming galaxies to define the boundaries between galaxy types \citep[e.g., Ke01,][]{2006MNRAS.371..972S}. In this paper we present grids of photoionization calculations that combine the AGN and starburst galaxy contributions (assuming either a constant or an instantaneous star-formation burst) to define new theoretical demarcation lines for mid-infrared and optical diagrams. Contrary to previous studies (i.e. in the optical), our new theoretical classification for starburst and AGN is based on the minimum and maximum AGN contributions to the optical emission line spectra as predicted from photoionization models of the mid-infrared diagrams.

\section{Theoretical  Models}

In order to investigate galaxy physical conditions in the mid-infrared and optical emission-line regions, we use constant density models with a hydrogen column density of $10^{21}{\rm cm^{-2}}$ throughout \citep[e.g.,][]{2010ApJ...725.2270P,2011ApJ...738....6M}, thus including radiation- and matter-bounded models depending on the ionization state \citep{2011ApJ...738....6M}. These considerations are to ensure a common set of conditions associated with the narrow-line region (NLR) illuminated by an AGN and with a star-forming region illuminated by a stellar radiation field.

We thus generate a grid of photoionization models varying the total hydrogen number density, $n_{\rm H}$, and the ionization parameter \cite[see][]{2006agna.book.....O},
\begin{equation}
U=\frac{1}{4\pi R^2cn_H}\int^\infty_{\nu_o}\frac{L_\nu}{h\nu}d\nu=\frac{Q(H)}{4\pi R^2cn_{\rm H}},
\label{u}
\end{equation}
where $R$ is the  distance to the cloud, $c$ is the speed of light and $Q(H)$ is the ionizing photon flux. All of our calculations have been carried out with the CLOUDY (version 10.00) photoionization code last described by \citet{2013RMxAA..49..137F}.

\subsection{The Active Galactic Nuclei}

We assume a broken power law to model the spectral energy distribution (SED) of an AGN, as used  by \citet{2011ApJ...738....6M} and similar to that suggested for NGC\,5548 and NGC\,4151 \citep{1998ApJ...499..719K,2000ApJ...531..278K}. It takes the form $F_\nu \propto \nu^{-\alpha}$ with $\alpha = 0.5$ below 13.6~eV, $\alpha =1.8$ from 13.6~eV to 1~keV and $\alpha = 0.8$ at higher energies. For the UV--Soft X-ray slope we take the median value derived from high-ionization mid-infrared emission lines for the BAT AGN \citep{2011ApJ...738....6M}, and the oxygen, neon and nitrogen abundances are assumed to be $-3.31$ \citep{2001ApJ...556L..63A}, $-4.0$ \citep{2001AIPC..598...23H} and $-4.07$ \citet{2001AIPC..598...23H}, respectively. Despite the fact that dust is likely to be mixed in with the emission-line gas in the Seyfert galaxy NLR   \citep[e.g.,][]{1986ApJ...307..478K,1993ApJ...404L..51N}, we do not include it in our AGN models because the dust-to-gas fraction is likely to vary within these regions \citep[e.g.,][]{1998ApJ...508..232K,2004ApJS..153...75G,2009ApJ...698..106K}. Therefore, to reduce the number of free parameters, we have adopted dust-free models for the NLR; however, at the end of Section~\ref{dust} we introduce dust in our final set of AGN models and discuss its effect.

\subsection{Starburst Galaxies}

Synthesis models from STARBURST99 are used to model star-forming galaxies that are optimized to reproduce the properties of galaxies with active star formation \citep{1999ApJS..123....3L,2005ApJ...621..695V,2010ApJS..189..309L}. We consider two cases for the star formation law: an instantaneous star-formation burst and star formation at a continuous rate. For the single stellar population or instantaneous star-formation burst, we adopt a total stellar mass of $1\times 10^6M_\odot$ to be distributed out between the upper and lower cut-off masses. Star-formation rate for the continuous models is 1~$M_\odot$$Yr^{-1}$. We also adopt a double power law for the stellar initial mass function (IMF) of the type presented by \citet{2001MNRAS.322..231K}. In this double Kroupa-type law there are two exponents, 1.3 and 2.3, for the low and high masses, respectively, that correspond to IMF intervals of 0.1--0.5 and 0.5--100 solar masses. For the stellar evolution model we use the Padova track with thermally pulsating asymptotic-giant-branch (AGB) stars. Three metallicities are considered in the present work, namely $Z = 0.050$, $Z=Z_\odot =0.020$ and $Z=0.004$. Finally, our models cover a stellar age range of 10$^6$--10$^8$ years with a time resolution of 0.1~Myr.

Based on these assumptions  we generate a grid of stellar spectral energy distributions. This stellar atmosphere grid in stellar age and metallicities can be compiled to use  as the input radiation source in CLOUDY to study the physical properties of the star formation regions in the galaxy. In order to model an H~II region we assume abundances to be the mean of those in the Orion Nebula as determined by \citet{1991ApJ...374..580B}, \citet{1991ApJ...374..564R}, \citet{1992ApJ...389..305O} and  \citet{1993ApJ...413..242R}. Similar to NLR modeling we assume  a single-zone, constant density model with hydrogen column densities of $10^{21}{\rm cm^{-2}}$. In addition, the H~II region is  characterized by a mixture of graphite and silicate grains typical of those along the Orion Nebula line of sight \citep{1991ApJ...374..580B}.

\section{Mid-Infrared Diagnostics}
\label{MidIRDiag}

Emission-line ratios can be used as a powerful tool to classify emission-line galaxies because they can provide insightful information about the source of ionizing radiation. As a starting point we focus our analysis on  mid-infrared diagnostics, in particular the emission-line diagnostic presented in W10 that utilizes three of the most common features observed in the mid-infrared spectra of emission-line galaxies, namely [Ne~II], [Ne~III] and [O~IV] lines.  As previously mentioned, the emission-line diagnostic diagram presented in W10 shows a clear separation between the different sources, with the majority of BAT AGN having [O~IV]/[Ne~III] and [Ne~III]/[Ne~II] ratios greater than unity; that is, AGN dominated.

\subsection{The Active Galactic Nuclei}

\subsubsection{Single-Zone Model}
\begin{figure}
\includegraphics[width=84mm]{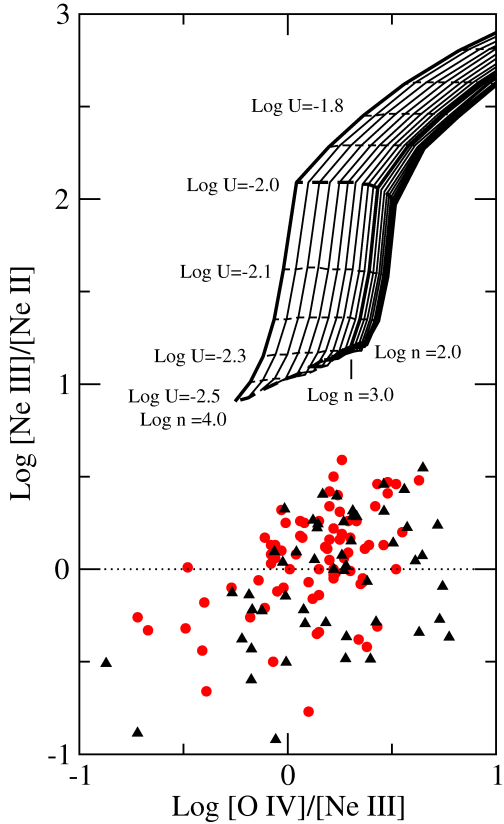}
  \caption{Comparison between the single-zone AGN photoionization model, the BAT AGN sample (red dots) and the  12$\umu$m sample of AGN \citep[black triangles, ][]{2010ApJ...709.1257T} overlaid on the W10 diagnostic diagram. On the grid, the solid lines represent density steps with $\Delta \log n_H= 0.1$ and the dashed lines ionization-parameter steps with $\Delta \log U= 0.1$. The dotted line represents the limit between AGN dominated,  [Ne~III]/[Ne~II] $>$ 1 \citep{2010ApJ...716.1151W}, and star-formation dominated sources.  \label{comp_agn}}
\end{figure}

Figure~\ref{comp_agn} shows a grid of single-zone photoionization  models overlaid on the W10 diagnostic for the BAT AGN sample. For the model grid presented in this comparison we use a range of densities and ionization parameters that are typical of the NLR \citep{2008ApJ...682...94M,2008ApJ...689...95M,2011ApJ...738....6M}. It is clear from Figure~\ref{comp_agn} that a single-zone  approximation misses the majority of BAT AGN by over-predicting the [Ne~III]/[Ne~II] ratios. This is in agreement with the fact that the [Ne~II] emission, given its low ionization potential,  may be originating in a different region than that producing the observed higher ionization [O~IV] and [Ne~III] emission \citep[e.g.,][]{2008ApJ...689...95M}.

\begin{figure}
  \includegraphics[width=84mm]{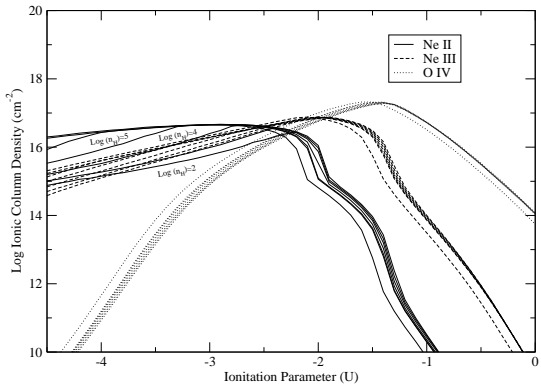}
  \caption{Predicted ionic column densities for Ne~II, Ne~III and O~IV as a function of ionization parameter $U$ and hydrogen densities in the range $n_{\rm H}= 10^{2}{-}10^{7}$~cm$^{-3}$ in steps of $\Delta \log n_{\rm H} = 1$.\label{ionic_columns} }
\end{figure}

In Figure~\ref{ionic_columns} we compare the predicted ionic column density and ionization parameter for Ne$^+$, Ne$^{++}$ and O$^{3+}$ to find the range of ionization parameters that can maximize  these mid-infrared emission lines. This comparison shows that [Ne~II] emission  peaks at a lower ionization parameter than [O~IV] and [Ne~III]. However, for $\log U < -2.5$, Ne$^+$ and Ne$^{++}$ can coexist in regions with densities of $\log n_{\rm H} \approx 10^4\,{\rm cm^{-3}}$ where O$^{3+}$ drops sharply. Therefore, in order to  simultaneously fit  the [Ne~III]/[Ne~II] and [O~IV]/[Ne~III] ratios of  the BAT AGN, we assume the physical conditions required to maximize [Ne~II] emission without increasing the predicted [Ne~III]; in other words, a region where only [Ne~II] is emitted efficiently. Figure~\ref{ionic_columns} also shows  that for densities $\log n_{\rm H} > 10^4\, {\rm cm^{-3}}$ there is a drastic change in the relative fraction between Ne$^+$ and Ne$^{++}$. In order to characterize the physical conditions in this region we fit  the observed [Ne~III]/[Ne~II] and [O~IV]/[Ne~III] ratios for the BAT AGN with a two-zone model.

\subsubsection{Two-Zone model}

In the two-zone approximation we expect to have a combination of matter- and radiation-bounded materials \citep[e.g.,][]{1996A&A...312..365B,2011ApJ...738....6M}. The matter-bounded component, where the material is optically thin to the ionizing radiation, corresponds  to the [O~IV] and [Ne~III] emitting region, whereas the radiation-bounded component is associated to the [Ne~II] efficient region where the material is optically thick. In order to constrain the density in the matter-bounded component we also reproduce the observed [Ne~V] 14.32$\umu$m/24.32$\umu$m ratio \citep[see][for values]{2010ApJ...716.1151W}. In \citet{2011ApJ...738....6M} it is found that the densities derived from the [Ne~V] ratio are similar to those of the [O~IV] emitting region, i.e. a fair assumption for our matter-bounded component.

\begin{figure}
  \includegraphics[width=84mm]{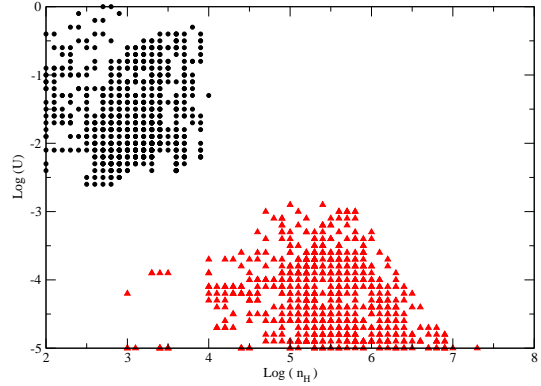}
  \caption{Parameter-space comparison for our two-zone AGN model. Circles characterize the high-ionization and low-density [Ne~III] and [O~IV] emitting region. Triangles characterize the low-ionization and high-density [Ne~II] emitting region.   \label{2param} }
\end{figure}

Figure~\ref{2param} shows the  parameter space that can reproduce simultaneously  the observed [Ne~III]/[Ne~II], [O~IV]/[Ne~III] and [Ne~V] ratios in  our two-zone model approximation for the BAT sources. It is clear from this comparison that there are two distinctive regions, a low-ionization and high-density region where [Ne~II] is efficiently emitted (dominated by radiation-bounded models) and a lower density but higher ionization state that corresponds to the [O~IV] and [Ne~III] region (dominated by matter-bounded models). These parameters are in good agreement with those presented in \citet{2008ApJ...689...95M} for an heterogeneous sample of Seyfert galaxies. According to this parameter-space distribution, the median values for the [Ne~II] dominated region are around $\log U =-4.0$ and $\log n_{\rm H}= 10^{5.5}~{\rm cm^{-3}}$, which are in excellent agreement with the maximization of Ne~II as implied from its ionic column density (see Figure~\ref{ionic_columns}).

\begin{figure*}
  \includegraphics[width=16.4cm]{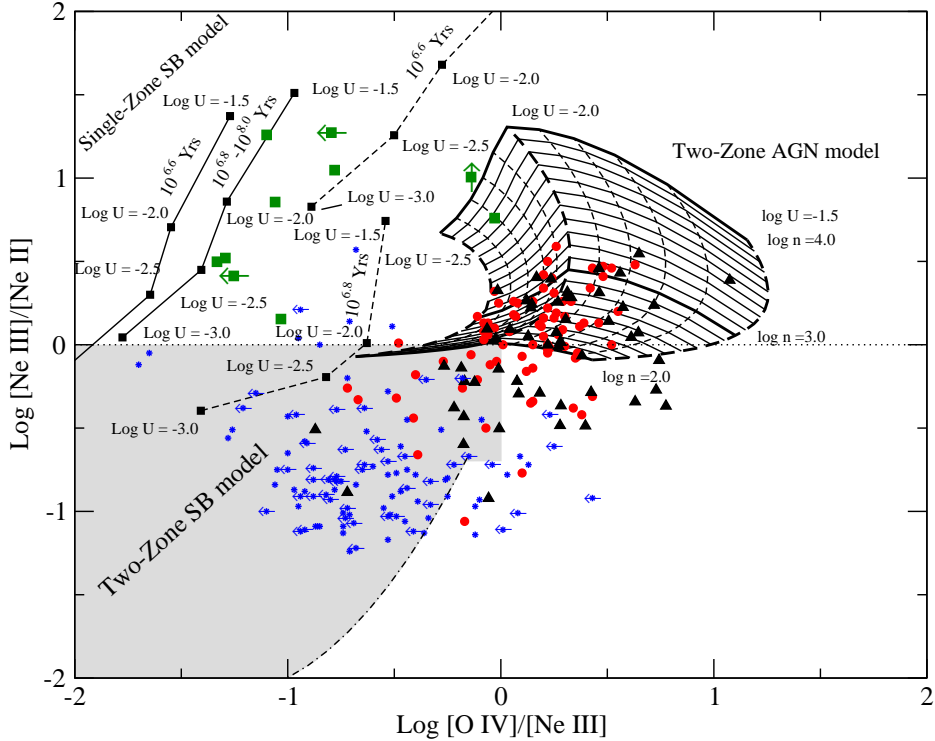}
  \caption{Comparison between different photoionization models and various observations over-plotted in the W10 diagnostic. We compare observations with three different photoionization models: 1) the two-zone AGN photoionization grid (solid and dashed black lines). On the grid, the solid lines represent density steps with $\Delta \log n_{\rm H}= 0.1$ and the dashed lines  ionization-parameter steps with $\Delta \log U= 0.1$, see text for details; 2) the single-zone starburst model for a metallicity of $Z=Z_\odot =0.02$ and a  fixed hydrogen density of $n_{\rm H}=10^{2.5}$~cm$^{-3}$. In this model, the solid lines represent a continuous star-formation law and the dashed lines an instantaneous star-formation burst and; 3) the  maximum prediction for the [Ne~III]/[Ne~II]  and  [O~IV]/[Ne~III] ratios based on a two-zone starburst model with solar metallicity (gray area). In this model, Equation~\ref{SB2} (dot-dashed line)    represents  the boundary of our two-zone SB models with solar metallicity. We compare these models with sources from different samples: the BAT AGN sample (red dots), the 12$\mu$m AGN sample (black triangles), the sample of BCD  from H09 (green squares) and,  H~II galaxies  from \citet{2010ApJ...725.2270P} (blue stars). The dotted line represents the limit between AGN dominated,  [Ne~III]/[Ne~II] $>$ 1, and the star-formation dominated sources. \label{weaver_solar}}
\end{figure*}

\begin{figure*}
  \includegraphics[width=16.4cm]{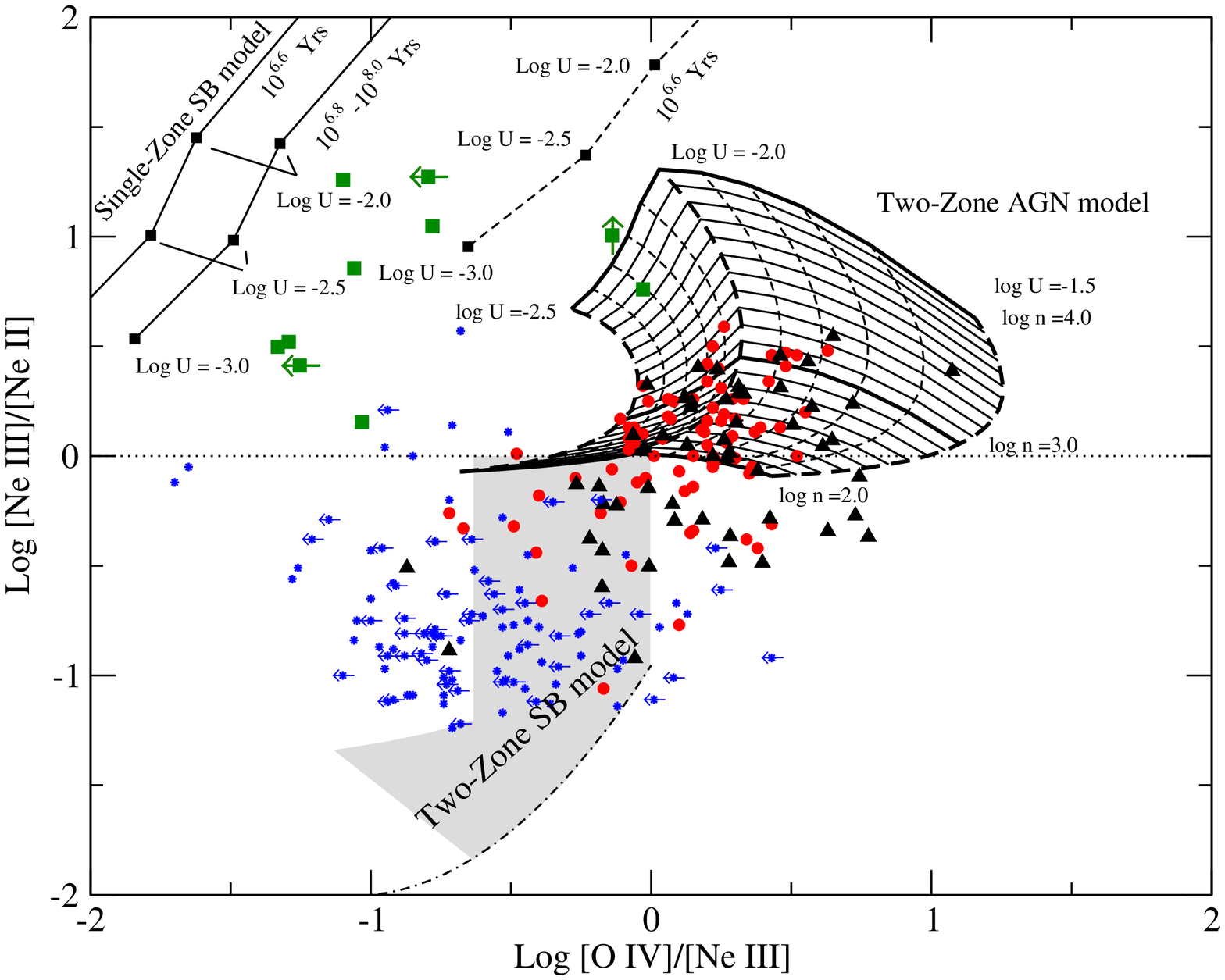}
  \caption{Comparison between different photoionization models and various observations over-plotted in the W10 diagnostic. We compare observations with three different photoionization models: 1) the two-zone AGN photoionization grid (solid and dashed black lines). On the grid, the solid lines represent density steps with $\Delta \log n_{\rm H}= 0.1$ and the dashed lines  ionization-parameter steps with $\Delta \log U= 0.1$, see text for details; 2) the single-zone starburst model for a metallicity of $Z = 0.004$ and a  fixed hydrogen density of $n_{\rm H}=10^{2.5}$~cm$^{-3}$. In this model, the solid lines represent a continuous star-formation law and the dashed lines an instantaneous star-formation burst and; 3) the  maximum prediction for the [Ne~III]/[Ne~II]  and  [O~IV]/[Ne~III] ratios based on a two-zone starburst model with subsolar metallicity, $Z = 0.004$ (gray area). In this model, Equation~\ref{SB2} (dot-dashed line)    represents  the boundary of our two-zone SB models with solar metallicity. We compare these models with sources from different samples: the BAT AGN sample (red dots), the 12$\mu$m AGN sample (black triangles), the sample of BCD  from H09 (green squares) and,  H~II galaxies  from \citet{2010ApJ...725.2270P} (blue stars). The dotted line represents the limit between AGN dominated,  [Ne~III]/[Ne~II] $>$ 1, and the star-formation dominated sources.  \label{weaver_subsolar}}
\end{figure*}

In Figure~\ref{weaver_solar} we present our two-zone model grid for the AGN. From the above discussion,  we select the median values for the [Ne~II] dominated region, namely $\log U =-4.0$ and $\log n_{\rm H}= 10^{5.5}~{\rm cm^{-3}}$, to be those of the radiation-bounded component. Then we select a range of density and ionization parameters for the [O~IV] and [Ne~III] component. The range of physical conditions for the latter (matter-bounded) are those derived from the individual fit of the mid-infrared ratios of the BAT sources (see Figure~\ref{2param}). Overall, our two-zone photoionization model is in much better agreement with observations; however, it is clear that several BAT AGN still lie outside our two-zone model. These sources, with [Ne~III]/[Ne~II] and [O~IV]/[Ne~III] ratios less than unity, have been described as galaxies dominated by star formation \citep[e.g.,][]{2008ApJ...689...95M,2010ApJ...716.1151W}. It must be noted that previous calculations only considered AGN ionization. Therefore these models will always under-predict the observed mid-infrared ratios in the presence of an additional  ionizing radiation field from star-formation regions, especially the [Ne~II] and [Ne~III] luminosities that are good tracers of stellar activity \citep[e.g.,][]{2007ApJ...658..314H}.  Additionally, Figure~\ref{weaver_solar} shows that our two-zone model has a lower limit for the [Ne~III]/[Ne~II] ratio around unity in agreement with observations of AGN dominated sources \citep[e.g.,][]{2008ApJ...689...95M,2010ApJ...716.1151W}.

\begin{figure}

  \includegraphics[width=84mm]{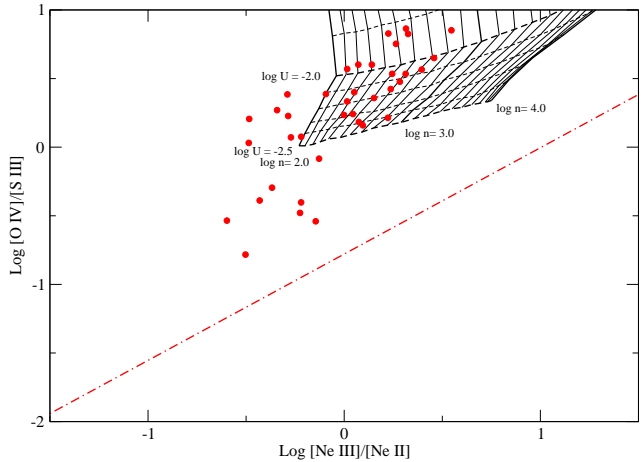}
  \caption{Comparison between our two-zone AGN photoionization models and the [O~IV]/[S~III] {\em vs.} [Ne~III]/[Ne~II] diagnostic presented in H09. The dot-dashed line represents the empirical separation between the AGN and the pure star-formation galaxy branch as defined by Equation~1 in H09. On the grid, the solid lines represent density steps with $\Delta \log n_{\rm H}= 0.1$ and the dashed lines ionization-parameter steps with $\Delta \log U= 0.1$. The red dots represent the Seyfert galaxies from the 12~$\umu$m sample \citep{2010ApJ...709.1257T} where we selected galaxies with detections in all four lines.
 \label{hao1} }
\end{figure}

We have generated a two-zone model grid that can successfully reproduce the AGN branch in W10. This analysis is to be extended to other mid-infrared diagnostics, in particular that presented in H09 involving the [Ne~III]/[Ne~II] and  [O~IV]/[S~III]33.4\micron~ ratios. The ionization energy to produce [S~III] (E$_{IP}$ = 23.33 eV) is similar to that for [Ne~II] emission, but the high density of the radiation-bounded component will suppress [S~III] via  collisional de-excitation.  On the other hand, given the critical density of the [S~III]33.4\micron~ transition (1.2$\times 10^3$ cm$^{-3}$), this line can be produced in the range of conditions of our matter-bounded component (i.e. the [Ne~III] and [O~IV] component), albeit with somewhat lower  ionization. Therefore, our two-zone model grid should be a good approximation to match the observed AGN branch in this diagnostic, specially at the lower end of the ionization range in the matter-bounded component. Figure~\ref{hao1} shows good agreement with observations, where it may be seen that our two-zone AGN models are located on the AGN branch different from pure starburst galaxies, probing the effectiveness of this diagnostic in separating emission line galaxies. For this comparison we only use values of Seyfert galaxies from the 12$\umu$m sample \citep{2010ApJ...709.1257T} where galaxies with detections in all four lines are selected. Note that there are no published values of [S~III] for the BAT sample. However, both samples have similar mid-infrared properties \citep[][]{2010ApJ...716.1151W}.   As discussed above, Figure~\ref{hao1} also shows that galaxies with a higher star formation contribution have lower [Ne~III]/[Ne~II] and  [O~IV]/[S~III]33.4\micron~ ratios, and fall outside of our two-zone AGN grid.

In general, in order to fit  AGN-dominated sources in the W10 diagram ([Ne~III]/[Ne~II] and [O~IV]/[Ne~III] ratios greater than unity), we require an AGN  [Ne~II] component. This high-density and low-ionization component is justified because of the existence of a number of sources with [Ne~II] emission that show no other signatures of star formation in their mid-infrared spectra; e.g. no detectable PAH features \citep[e.g.,][]{2007ApJ...671..124D,2010ApJ...709.1257T,2012ApJ...758....1L,2014MNRAS.441..630S} suggesting that there must be some [Ne~II] emission directly associated to the AGN. As seen in Figure~\ref{weaver_solar} this component, where [Ne~II] is emitted efficiently, cannot explain the observed  emission in many objects where star formation is the dominant ionizing source. As we will discuss below, ionization from stars is more efficient in producing [Ne~II] than the harder AGN radiation field \citep[e.g.,][]{2008ApJ...689...95M,2012ApJ...758....1L}. Therefore, our [Ne~II] AGN component is chosen to represent the maximum (possible) AGN-related [Ne~II] emission, but when compared to the total observed flux, this component may be a minimum to the observed [Ne~II] emission.  It is important to emphasize that, despite the fact that we fix the parameters of the radiation-bounded component to the median of  those that can reproduce simultaneously the observed [Ne~III]/[Ne~II], [O~IV]/[Ne~III] and [Ne~V] ratios, the ionic column density of Ne~II is basically insensitive (constant) to  changes around these  values (see Figure~\ref{ionic_columns}), suggesting that our two-zone grid is stable to  small differences in the  conditions that we selected for the  [Ne~II] AGN component. 

\subsection{Starburst Galaxies}
\subsubsection{Single-Zone Model}

\begin{figure}
  \includegraphics[width=74mm,angle=-90]{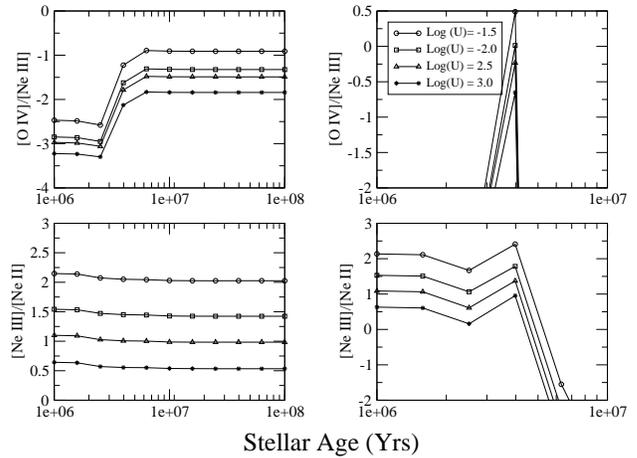}
  \caption{Comparison of mid-infrared ratios in the W10 diagram for a metallicity of $Z = 0.004$, a fixed hydrogen density of $n_{\rm H}=10^{2.5}$~cm$^{-3}$ and different ionization parameters. Left panels: continuous star-formation law. Right panel: instantaneous star-formation burst. \label{ratios1} }
\end{figure}

\begin{figure}
  \includegraphics[width=74mm,angle=-90]{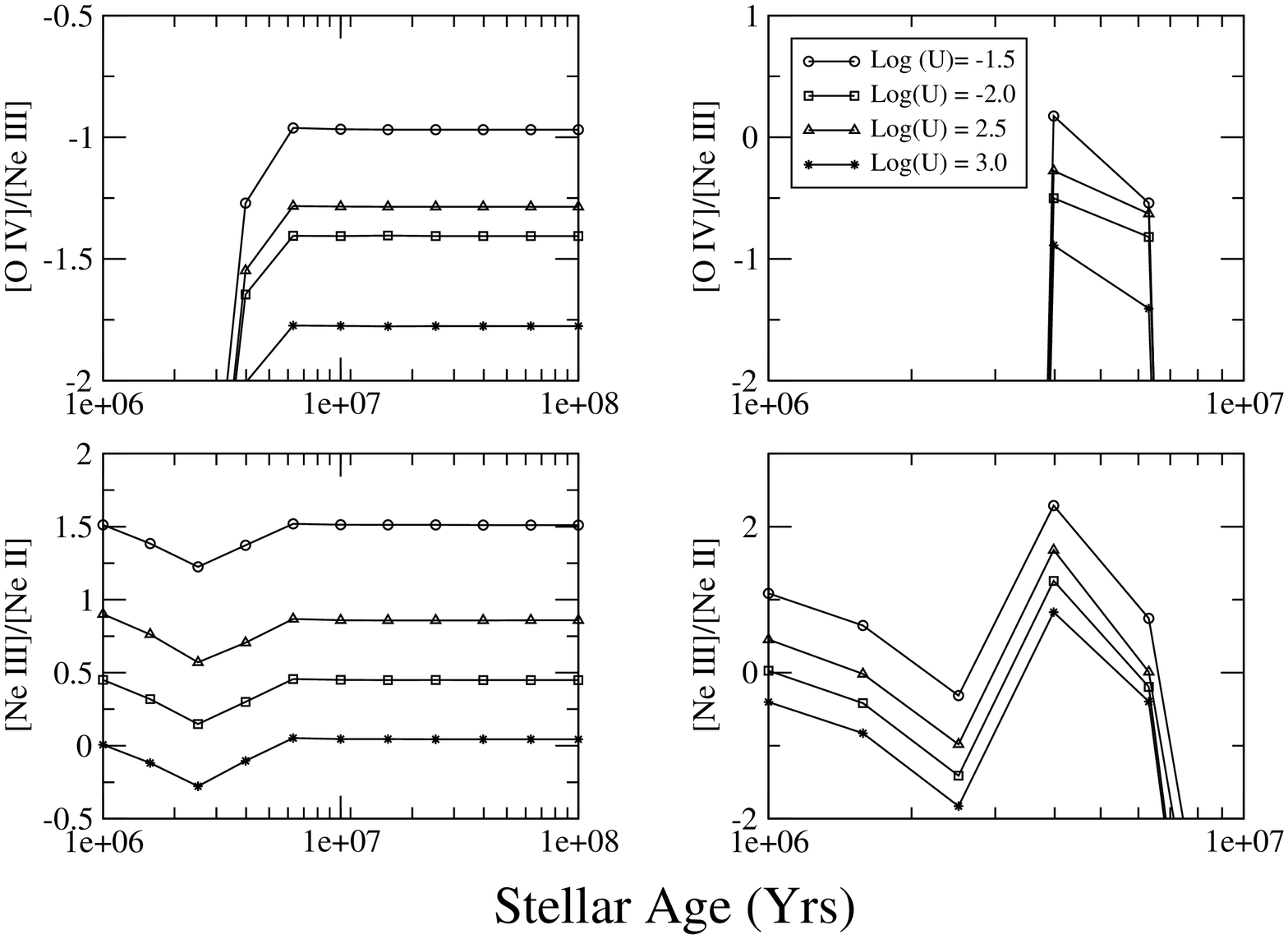}
  \caption{Comparison of mid-infrared ratios in the W10 diagram for a metallicity  of $Z=Z_\odot= 0.020$, a fixed hydrogen density of $n_{\rm H}=10^{2.5}~{\rm cm^{-3}}$ and different ionization parameters. Left panels: continuous star-formation law. Right panel: instantaneous star-formation burst.\label{ratios2} }
\end{figure}

\begin{figure}
  \includegraphics[width=74mm,angle=-90]{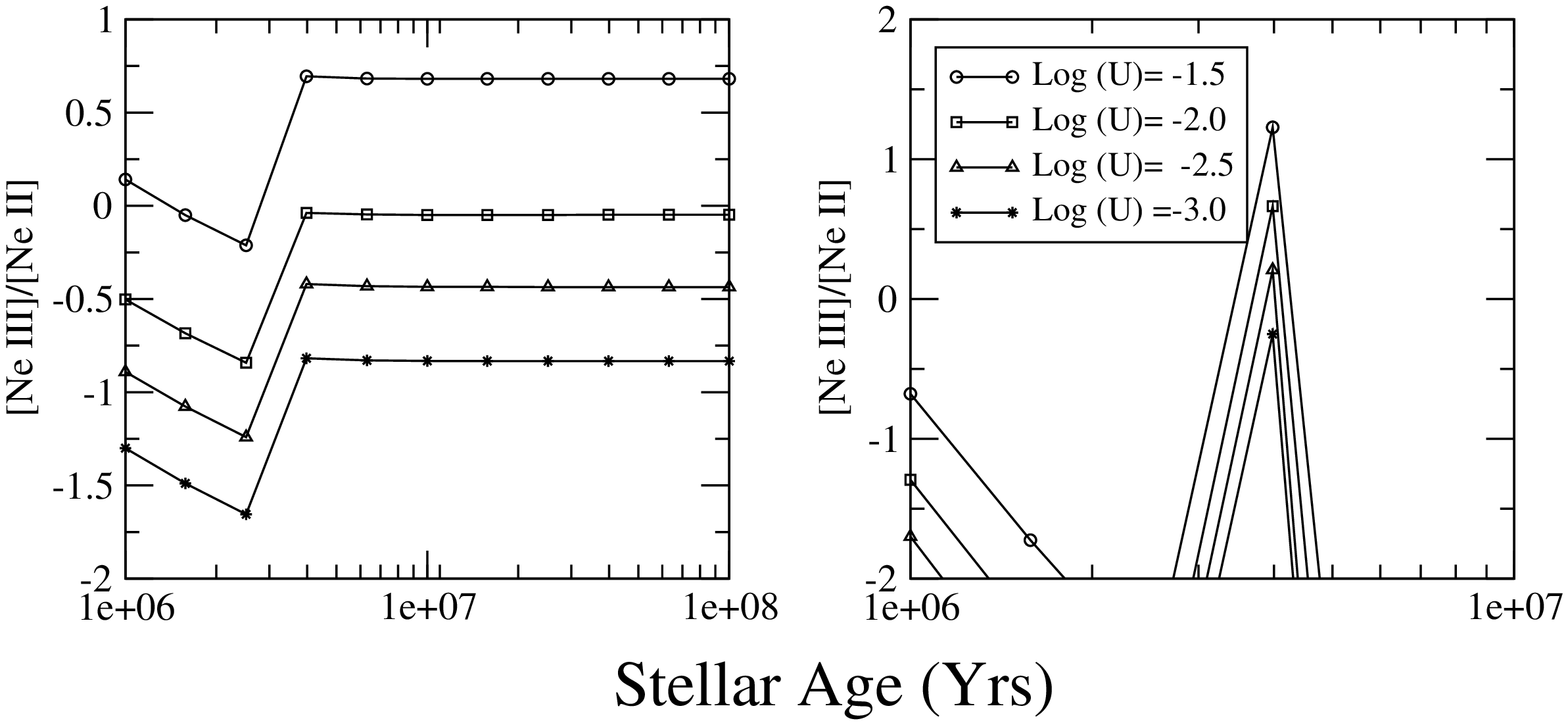}
  \caption{Comparison of mid-infrared ratios in the W10 diagram for a metallicity  of  $Z=0.05$, a fixed hydrogen density of $n_{\rm H}=10^{2.5}$~cm$^{-3}$ and different ionization parameters. Left panels: continuous star-formation law. Right panel: instantaneous star-formation burst.\label{ratios3} }
\end{figure}

Figures~\ref{ratios1}--\ref{ratios3} show  a comparison of our models  for  each mid-infrared ratio presented in the W10 diagram   as a function of stellar age, metallicity and  star-formation law for a set of conditions typical of H~II regions: $-3.0 < \log U < -1.5$ and  $n_{\rm H} \sim 10^{2.5}~{\rm cm^{-3}}$ \citep[e.g.,][]{2004ApJ...606..237R} with a fixed  hydrogen density of $n_{\rm H}=10^{2.5}~{\rm cm^{-3}}$ and a set of four ionization parameters.

\begin{figure}

  \includegraphics[width=84mm]{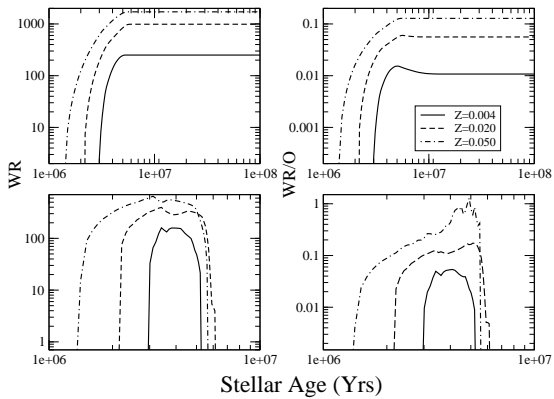}
  \caption{Number of WR stars and ratio of WR to O stars as a function of time for different metallicities of the evolutionary model. {\it Upper panel}: continuous star-formation law. {\it Lower panel}: star-formation burst. \label{WR}}
\end{figure}

For both star-formation laws there are dramatic changes in the contribution of stellar populations to the mid-infrared emission lines. Ratios with emission lines that require energies $E>40$~eV, e.g. [Ne~III] and  [O~IV], have their strongest emission at $\sim 4$~Myr during the  Wolf-Rayet  phase when  WR stars contribute the most to the far-UV flux. After a short time period the  burst population becomes very weak at $\ga 6.5$~Myr.  For the continuous star formation case after $\sim 6.5$~Myr the stellar population reaches an equilibrium between birth and death (of massive stars) thus becoming time independent. On the other hand,  lower-ionization lines such as [Ne~II] can be related to the presence of massive main-sequence O stars where they have their biggest contribution at $< 3$~Myr. However, [Ne~II] remains strong even at 10~Myr. Interestingly, [Ne~II] emission fades away during a short time period,  around  $\sim 4$~Myr during  the WR phase, because  the harder photoionizing continuum from WR stars  ionizes Ne$^+$  into Ne$^{2+}$. It must be pointed out that the WR phase is very metallicity dependent. Figure~\ref{WR} shows a comparison for the instantaneous and continuous star formation case for the number of WR stars and the WR/O ratio from the models as a function of stellar age. As it can be seen from these comparisons, the number of WR stars and the WR/O ratio is highly sensitive to metallicity with  a higher prediction of WR stars and longer WR phase in starburst with higher metallicities \citep[e.g., ][]{1999ApJS..123....3L}.

In Figures~\ref{ratios1}--\ref{ratios3} it is important to note that the [O~IV]/[Ne~III] ratio can only be effectively reproduced after massive O stars have evolved into their WR phase; in other words, according to our synthesis models WR stars are the only source of ionizing photons capable of producing  the  [O~IV] and [Ne~V] emission observed in non-AGN sources. Figures~\ref{weaver_solar}--\ref{weaver_subsolar} show a  comparison between  our  single-zone starburst photoionization models  and  a sample of extreme starbursts, blue compact dwarf galaxies presented in H09 over-plotted in the W10 diagram. This model grid covers the range of ionization parameters $-3.0\leq \log U \leq -1.5$ and a fix density of $\log n_{\rm H}=10^{2.5}$. Our single-zone model grid with solar metallicity for a star-formation bursts shows the best agreement with the observation of BCD from H09, where  many of the galaxies also show optical signatures typical of WR galaxies in agreement with our stellar models and the previous discussion. At subsolar metallicities our models can reach higher [O~IV]/[Ne~III] values in closer agreement with the observed ratio of the low-metallicity galaxy IZw18, [O~IV]/[Ne~III] = 0.93 (H09). It must be emphasised that we are not using our grid of photoionization models to predict the emission of individual galaxies but rather to investigate the best range in parameter space for each region in the W10 diagram.

\subsubsection{Two-Zone Model}

Our single-zone models for an H~II region can successfully reproduce the observed  [O~IV]/[Ne~III]~$<$ 1 and [Ne~III]/[Ne~II]~$>$ 1 ratios by invoking typical density values and ionization parameters found in H~II regions. These ratios in  W10 correspond to extreme starburst galaxies such as the one presented in H09 that comprise BCD galaxies, many of which show the optical signatures of WR galaxies. However, our single-zone model fails to predict galaxies with [Ne~III]/[Ne~II]~$<$ 1 and  [O~IV]/[Ne~III]~$<$ 1, a region that is characteristic of  starburst and H~II galaxies. As previously discussed and according to our models, only WR stars have an ionizing continuum capable of producing O$^{+3}$; therefore, in order to fit the observations of starburst and H~II galaxies in this region, a different scenario must be considered.

A possible scenario is the inclusion of an extra ionizing component.  In order to reproduce  the observed ratios for the H~II and starbust galaxies in the W10 diagram, our two-zone model needs to simultaneously maximize [Ne~II] and [O~IV] emissions. It is worth noting that our two-zone model for star-forming galaxies has many free parameters, e.g. star formation law, metallicity, stellar age, ionization parameter and hydrogen density for each component. Therefore, caution must be taken as the following  analysis is based on a set of assumptions that will  maximize the amount of [Ne~II] and [O~IV] emission so as to fit observations for galaxies with [Ne~III]/[Ne~II]~$<$ 1 and  [O~IV]/[Ne~III]~$<$ 1 where our single-zone model is inadequate.

Figure~\ref{weaver_solar} shows a comparison between  H~II galaxies \citep{2010ApJ...725.2270P} and the  predicted region  for the [Ne~III]/[Ne~II]  and  [O~IV]/[Ne~III] ratios based on our two-zone starburst  model with solar metallicity. For the  two-zone model we use  any  combination of an  instantaneous star-formation burst within a range of ionization parameters of $-3.0 < \log U < -1.5$  for each zone. The gray region in this comparison represents the location of  all the   predicted ratios with values  [Ne~III]/[Ne~II]~$<$ 1 and  [O~IV]/[Ne~III]~$<$ 1 from our two-zone starburst approximation.  These are the ratios of the location for the SB/H~II galaxies in the W10 diagram. Note that each component in our two-zone model  also has a range of stellar age (10$^6$--10$^8$ years) and hydrogen density. From the previous considerations, we perform the quadratic fit 
\begin{equation}
\begin{split}
\log {\rm [Ne~III]/ [Ne~II]} =  -0.27 + 2.79\times \left(\log {\rm [O~IV]/ [Ne~III]}\right)\\
 + 1.06 \times \left(\log {\rm [O~IV]/ [Ne~III]}\right)^2
\label{SB2}
\end{split}
\end{equation}
to the boundary of the maximum contribution from our two-zone model to the W10 diagram.

Equation~\ref{SB2} represents the maximum contribution possible to the W10 diagram derived from our two-zone starburst  models with solar metallicity. In other words, it  represents the boundary of our two-zone starbust  models. Interestingly, Figure~\ref{weaver_solar} shows  a few observed sources with ratios outside  the previous equation, i.e. outside  the  starbust region. These sources are NGC\,253, NGC\,3949, NGC\,4157, NGC\,520, NGC\,5907 and NGC\,660. As previously discussed, emission from highly ionized species such as O$^{3+}$ and Ne$^{4+}$ cannot be easily produced by stars because they require a hard photoionizing continuum that can only be produced at a specific period during the WR phase of massive main-sequence stars. Moreover, the duration of the WR phase in these massive stars  is highly sensitive to the metallicity of the evolutionary models. Therefore, for galaxies outside our W10 demarcation line  one must consider other possibilities such as a weak AGN instead of a population of WR stars  \citep[e.g., ][]{1998A&A...333L..75L} or  excitation to high-ionization lines by shocks \citep[e.g.,][]{2004ApJS..153....9G}. Interestingly, some of these sources show evidence for a low-luminosity AGN, e.g. NGC\,660 \citep{2006A&A...455..773V} and NGC\,253 \citep{2010ApJ...716.1166M}.

A similar analysis can be carried out with a combination of an instantaneous star-formation burst with subsolar metallicity. Figure~\ref{weaver_subsolar} shows the  predicted region  for the [Ne~III]/[Ne~II]  and  [O~IV]/[Ne~III] ratios based on our  two-zone starburst  model with subsolar metallicity ($Z=0.004$). Again, we perform a quadratic fit to the boundaries of our two-zone model:
\begin{equation}
\begin{split}
\log {\rm [Ne~III]/ [Ne~II]} =  -0.95 + 1.98\times \left(\log {\rm [O~IV]/ [Ne~III]}\right)\\ 
 + 0.93 \times \left(\log {\rm [O~IV]/ [Ne~III]}\right)^2\ .
\label{SB3}
\end{split}
\end{equation}
Equation~\ref{SB3} represents  the boundary of our two-zone starbust models with subsolar metallicity.

It is important to mention that any model combination with a continuous star formation law is unable to reach mid-infrared ratios [O~IV]/[Ne~III] $\ga  0.1$ as seen in Figures~\ref{weaver_solar}--\ref{weaver_subsolar}; thus a combination of continuous star formation law models will provide a poor representation to the observed values of starburst and H~II galaxies in the W10 diagram. On the other hand, Figures~\ref{weaver_solar}--\ref{weaver_subsolar} show that a combination of AGN and SB models can  also reproduce most of the observed line ratios.

\section{Optical Diagnostics}

In Section~\ref{MidIRDiag} we have discussed various photoionization models that can successfully reproduce emission-line galaxies in different mid-infrared diagnostics, ranging from pure starburst to AGN.  In this respect,  we use our  mid-infrared models to predict some of the most commonly used  emission-line ratios in optical classification diagrams, e.g. [O~III]/H$\beta$ and [N~II]/H$\alpha$, to establish new classification parameters. For the present work we have compared our theoretical predictions with a sample of galaxies from SDSS DR8. The galaxy spectra for  DR8 refer to the measurements and techniques developed at the  Max Planck Institute for Astrophysics and the Johns Hopkins University \citep[the MPA-JHU measurements are based on the methods of][]{2004MNRAS.351.1151B,2004ApJ...613..898T}. Our SDSS sample was selected to have a S/N$>$ 3$\sigma$ for the four emission lines in the [O~III]/H$\beta$ and [N~II]/H$\alpha$ ratios resulting in a total of 411,010 galaxies.

\subsection{New Classification Parameters\label{dust}}

\begin{figure}

  \includegraphics[width=84mm]{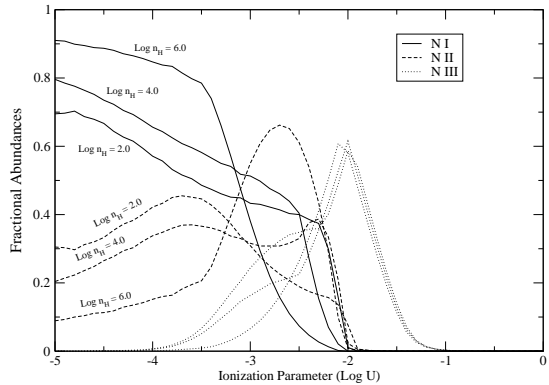}
  \caption{Radius-averaged ionization fractions for N~I, N~II and N~III for different densities as a function of ionization parameter.\label{nitrogen} }
\end{figure}

The origin of N$^+$ emission, giving its low ionization potential ($E_{\rm IP} = 14.53$~eV), can be associated to the same source of ionizing radiation as Ne$^+$ emission; however, the [N~II]$\lambda$6583 (hereafter [N~II]) line has a critical density ($6.6\times 10^4$~cm$^{-3}$) roughly one order of magnitude lower than [Ne~II] ($7.1\times 10^5$~cm$^{-3}$). Therefore, in our  two-zone AGN approximation, the  high-density radiation-bounded component where the [Ne~II] is emitted efficiently  represents a conservative  lower limit for [N~II] emission. Figure~\ref{nitrogen} shows the radius-averaged ionization fraction for N$^0$, N$^+$ and N$^{++}$ for a range of densities and ionization parameters, where it can  be seen that, for the set of conditions that maximize [Ne~II] emission (Figure~\ref{2param}), most of the nitrogen is neutral and only a small fraction ($\sim 20\%$) is singly ionized. Additionally, the high density in the radiation-bounded component is suppressing [N~II] emission via collisional de-excitation. Therefore, it is clear that the physical conditions in the [Ne~II] emitting region are not optimal for producing [N~II]; however, even under these unfavorable conditions, a small fraction of N$^+$ can exist. As a result, our two-zone grid is not intended to fit the entire AGN branch in the optical [O~III]/H$\beta$ and [N~II]/H$\alpha$ diagram, but to provide a reliable lower-limit to the AGN contribution in these optical emission lines. Moreover, the stability  of this lower-limit relies on the fact that  the predictions from the [Ne~II] AGN component are nearly independent  to small changes around the values that we set for the radiation-bounded component. 

\begin{figure}
  \includegraphics[width=84mm]{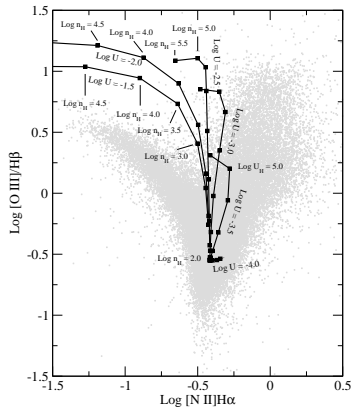}
  \caption{Comparison between the SDSS sample and the predicted optical emission lines derived from our two-zone mid-infrared AGN models. Similarly to Figure~\ref{weaver_solar}, we fix the parameters from one component to maximize [Ne~II] emission, namely $\log U =-4.0$ and $\log n_{\rm H}= 10^{5.5}$~cm$^{-3}$, and span ranges of density and ionization parameter for the [O~IV] and [Ne~III] component\label{sdss} }
\end{figure}

\begin{figure}
  \includegraphics[width=84mm]{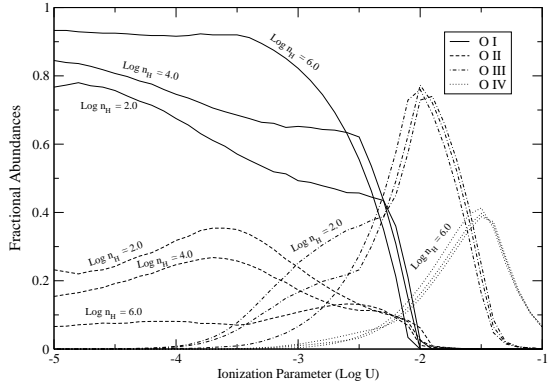}
  \caption{Radius-averaged ionization fraction for O~I, O~II, O~III and O~IV for different densities as a function of ionization parameter.\label{oxygen} }
\end{figure}

Figure~\ref{sdss} shows the optical predictions from our two-zone AGN approximation (see Figure~\ref{weaver_solar}) overlaid on our SDSS sample for the [O~III]/H$\beta$ {\em vs.} [N~II]/H$\alpha$ diagram. It may be appreciated that our two-zone grid under-predicts the observed [N~II]/H$\alpha$ as previously discussed.  On the other hand, [O~III] emission can  originate in a region with very similar conditions to [Ne~III] and [O~IV] emission \citep[e.g.,][]{2008ApJ...682...94M}. The physical conditions of the [Ne~III] and [O~IV] emitting region  are in good agreement with those inferred by \citet{2005MNRAS.358.1043B} from the simultaneous fit of the equivalent width of the narrow [O~III], [O~III]$\lambda$4363 and H$\beta$ lines in the Palomar-Green quasar sample; in particular to those conditions in the outer region in their two-zone approximation, where most of the [O~III] is emitted. Therefore, our matter-bounded component   can efficiently produce the observed  [O~III]/H$\beta$. We can then identify a lower-bounded demarcation line for our models corresponding to a constant ionization parameter of $\log U = -1.5$. This value for the ionization parameter is where the radius-averaged ionization fraction of O$^{+2}$ decreases as this species gets ionized into O$^{+3}$ (see Figure~\ref{oxygen}). In other words, this lower-bounded demarcation line represents an ionization parameter where the O$^{3+}$ surpasses the ionization fraction of O$^{2+}$; therefore, the gas is inefficient in producing [O~III] emission, thus,  a reasonable limit for our models. In addition, $\log U = -1.5$ is the limit found by \citet{2005MNRAS.358.1043B} in the distribution of ionization parameters for the outer region in their two-zone approximation. From a nonlinear fit of the two-zone photoionization model with a fixed  $\log U = -1.5$, we define a galaxy to have an AGN component if
\begin{equation}
\log {\rm [O~III]/H_\beta} > \frac{0.12}{\left ( \log {\rm [N~II]/H_\alpha} +0.34\right )}+1.15\ .
\label{eq_sdss}
\end{equation}

\begin{figure}
  \includegraphics[width=84mm]{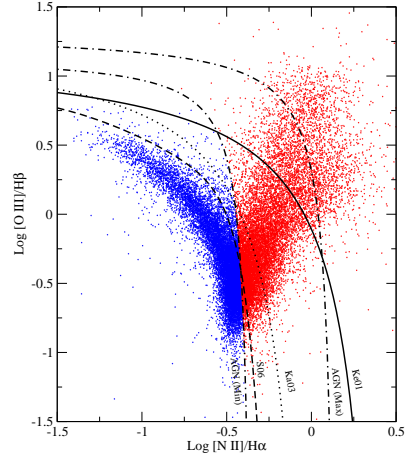}
  \caption{Comparison between different demarcation lines to separate AGN and star-forming galaxies in the [O~III]/H$\beta$ {\em vs.} [N~II]/H$\alpha$ diagram. The dot-dashed lines are demarcations derived from the nonlinear fit to the boundary of our two-zone AGN dominated model that define the minimum (AGN min.) and maximum (AGN max.) AGN conditions (Equation~\ref{eq_sdss} and Equation~\ref{eq_sdss1}, respectively). The dotted line is the empirical demarcation presented in \citet{2003MNRAS.346.1055K} (Ka03). The solid line is the theoretical  demarcation line for the maximum contribution calculated  in \citet{2001ApJ...556..121K} (Ke01). Finally, the dashed line is the theoretical line for the pure star-forming galaxies from \citet{2006MNRAS.371..972S} (S06). Star-forming galaxies are shown in blue, AGN-dominated sources in red according to our demarcation line, Equation~\ref{eq_sdss}. \label{sdss_lines} }
\end{figure}

This demarcation line is the lowest theoretical prediction for the AGN contribution to the optical spectra, and lies above the star-formation branch of the [O~III]/H$\beta$ {\em vs.} [N~II]/H$\alpha$ diagram predicting a lower number of pure star-forming galaxies than Ka03. In fact, the Ka03 demarcation line separates our SDSS sample into $\sim$65$\%$ pure star-forming galaxies (266,696 sources) and $\sim$35$\%$ AGN (144,314 sources), while our definition leads to a separation of the sample of    $\sim$51$\%$ pure star-forming galaxies (208,977 sources) and $\sim$49$\%$ AGN (202,033 sources). Note that in our AGN dominated models any extra contribution from star-forming regions can only move objects above the AGN lower-bound demarcation line (Ke01) as we push into more favorable conditions for [N~II]. Therefore, sources above this line may have different stellar contributions to their optical spectra, ranging from composite systems to AGN dominated sources. Figure~\ref{sdss_lines} shows a comparison between our new AGN demarcation line and those from Ke01 and Ka03. Overall, our AGN models predict a higher AGN population by a factor of $\sim$ 1.4 than predictions derived from a semi-empirical division of star-forming galaxies (K03).

The expression for the boundary between star-forming galaxies and AGN, namely Equation~\ref{eq_sdss}, is in agreement  with the theoretical  curve presented by \citet{2006MNRAS.371..972S}, which is the demarcation between ``pure" star-forming galaxies and galaxies hosting an AGN. This line represents the  set of stellar models that best reproduce the upper envelope of the  left wing in the [O~III]/H$\beta$ {\em vs.} [N~II]/H$\alpha$ diagram. Using this curve, our SDSS sample is divided into  $\sim$49$\%$ pure star-forming galaxies (203,156 sources) and $\sim$51$\%$ AGN (207,854 sources), similar to the cut from our line. Interestingly, our curve lies above that from \citet{2006MNRAS.371..972S} with a large gap above the left wing (star-formation branch) in the optical diagram. It is important to point out that the predicted ionization fraction for highly ionized ions is density independent in the optically thin, matter-bounded component. On the other hand, the ionization fraction in our optically thick, radiation-bounded models with more lowly ionized ions is very density sensitive (see Figures~\ref{nitrogen} and \ref{oxygen}).

We can define a theoretical limit for the maximum AGN contribution to the optical diagram based on a two-zone approximation. For this purpose, we simultaneously fit the observed [O~III]/H$\beta$ and [N~II]/H$\alpha$ ratios for each AGN in our SDSS DR8 sample with a two-zone model grid spanning ranges of the ionization-parameter ($-3.5\leq \log U \leq -1.5$) and hydrogen density ($2.0\leq \log n_H \leq 5.5$~cm$^{-3}$) for each component. These parameter ranges are typical of the AGN NLR, including the conditions driving [O~III] emission \citep[e.g.,][]{2005MNRAS.358.1043B,2008ApJ...682...94M,2008ApJ...689...95M,2010ApJ...725.2270P}, and moreover, are consistent with the set of conditions derived from the ionization fraction for these lines (see Figure~\ref{nitrogen} and Figure~\ref{oxygen}). Figure~\ref{sdss_lines} also shows the result of a nonlinear fit to the upper-bound region derived from our models from which we define the maximum AGN line
\begin{equation}
\log {\rm [O~III]/H_\beta} = \frac{0.16}{\left ( \log {\rm [N~II]/H_\alpha} -0.17\right )}+1.31
\label{eq_sdss1}
\end{equation}

This line represents the maximum AGN contribution to the optical spectra given by the set of conditions adopted in our AGN models, but it is clear that a number of AGN lie above it. To reach this extreme upper-right region in the AGN branch our models need to increase the predicted [O~III]/H$\beta$ and [N~II]/H$\alpha$ ratios. A possible way is to increase the oxygen and nitrogen abundances taking into account that at high-metallicity they scale proportionally \citep[e.g.,][]{2004ApJS..153....9G}. An alternative way would be to increase the hydrogen column density in the models resulting in a combination of radiation-bounded regions that will increase [O~III] and [N~II] production \citep[e.g.,][]{2011ApJ...738....6M}. Given the number of free parameters, we are unable to provide a unique scenario to explain the AGN-branch high end in the  [O~III]/H$\beta$ and [N~II]/H$\alpha$  diagram above our AGN maximum demarcation line and outside our parameter space.

\begin{figure}
  \includegraphics[width=84mm]{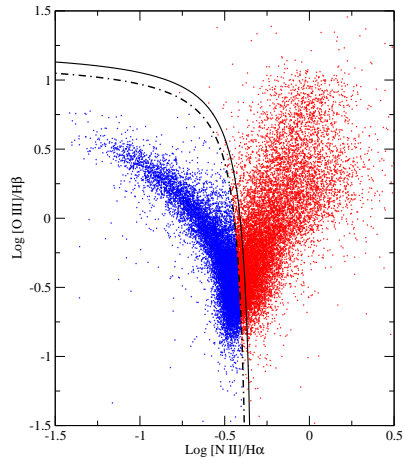}
  \caption{Comparison between different demarcation lines to separate AGN and star-forming galaxies in the [O~III]/H$\beta$ {\em vs.} [N~II]/H$\alpha$ diagram. The dot-dashed line is the demarcation derived from the nonlinear fit to the boundary of our two-zone AGN dominated model (dust free, Equation~\ref{eq_sdss}). The solid line is the nonlinear fit to the boundary of our two-zone AGN dominated model with the inclusion of dust, assuming a dust-to-gas ratio of 50\% of the ISM value  (Equation~\ref{eq_sdss_dust}). Star-forming galaxies are shown in blue, AGN-dominated sources in red according to our demarcation line, Equation~\ref{eq_sdss}.  \label{sdss_lines_dust} }
\end{figure}

As previously mentioned, our AGN photoionization models are dust free; however, to study the effect of dust on our demarcation lines, we generate a new set of models where grains mix with the emission-line gas. We include graphite and silicate grains  with a size distribution and a dust-to-gas ratio half of the value for the Galactic interstellar medium (ISM) \citep{1977ApJ...217..425M}. This dust-to-gas ratio is based on the high angular resolution analysis of the {\it Hubble Space Telescope} (HST)/{\it Space Telescope Imaging Spectrograph} (STIS) of the  NGC\,4151 and NGC\,1068 Seyfert galaxies \citep[][]{2000ApJ...531..278K,2000ApJ...544..763K,2000ApJ...532..256K}. From these models we have derived
\begin{equation}
\log {\rm [O~III]/H_\beta} > \frac{0.13}{\left ( \log {\rm [N~II]/H_\alpha} +0.31\right )}+1.24\ .
\label{eq_sdss_dust}
\end{equation}
Equation~\ref{eq_sdss_dust} represents the AGN demarcation line including dust and the same set of physical conditions derived from our dust-free models  (Equation~\ref{eq_sdss}). Figure~\ref{sdss_lines_dust} gives a demarcation line comparison with and without dust, where it may be seen that dust has a small effect on our theoretical line in the [O~III]/H$\beta$ {\em vs.} [N~II]/H$\alpha$ diagnostic diagram.

\begin{figure}
  \includegraphics[width=84mm]{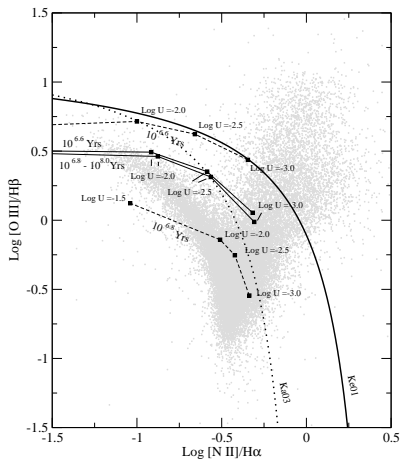}
  \caption{Comparison between the SDSS sample and the predicted optical emission lines derived from our single-zone mid-infrared SB models. The SB models  presented here are the same as in  Figure~\ref{weaver_solar}. The dashed and solid lines represent the same stellar law as in Figure~\ref{weaver_solar}. The dotted line is the empirical demarcation presented in \citet{2003MNRAS.346.1055K} (Ka03). The solid line is the theoretical  demarcation line for the maximum contribution calculated  in \citet{2001ApJ...556..121K} (Ke01)\label{comp_sdss_SB} }
\end{figure}

Similarly, in Figure~\ref{comp_sdss_SB} we overlay our single-zone starburst optical predictions with solar metallicities on  our SDSS sample in the [O~III]/H$\beta$ {\em vs.} [N~II]/H$\alpha$ diagram. From this comparison it is found that our single-zone models for starburst galaxies are a good match to the pure star-formation branch of galaxies with the ionization parameter decreasing with metallicity toward low [N~II]/H$\alpha$ and high [O~III]/H$\beta$. Our results are in excellent agreement with previous estimates based on the theoretical modeling  of the maximum starburst contribution line presented in Ke01 and the empirical demarcation line presented in Ka03 between pure SB/H~II galaxies and sources with some AGN contribution to their optical spectra. As previously discussed, our two-zone starbust  models are extremely sensitive to a number of parameters that can lead to some degeneracy; for this reason and for the sake of simplicity, we do not show a comparison between our two-zone predictions and the SDSS sample.

\section{Conclusions}

We have carried out extensive and detailed photoionization modeling to successfully locate different emission lines galaxies in optical and mid-infrared diagnostic diagrams. Our model grids cover a wide range in parameter space for the non-thermal AGN continuum and starburst galaxies with different stellar population laws and metallicities. As a result we arrive at the following findings.
\begin{enumerate}
  \item We successfully reproduce the AGN region in the W10 diagram with a two-zone model: [O~IV]/[Ne~III]$>1$ and [Ne~III]/[Ne~II] $>1$. This model is a combination of radiation-bounded (high density and low ionization) and  matter-bounded (low density and high ionization) components that efficiently produce [Ne~II] and [Ne~III]--[O~IV] emissions, respectively.

  \item Overall, our single-zone starburst models are a good fit to the extreme starburst region in the W10 diagram, i.e. [O~IV]/[Ne~III]$<1$ and [Ne~III]/[Ne~II] $>1$,  for which the instantaneous star-formation burst provides the best fit. In addition, our subsolar models are able to reach the highest [O~IV]/[Ne~III] values in agreement with the location of low-metallicity starburst galaxies.

  \item The  two-zone starburst approximation for an instantaneous star-formation burst provides the best fit to the star-forming region in the W10 diagram, i.e. [O~IV]/[Ne~III]$<1$ and [Ne~III]/[Ne~II] $<1$. However, we find that any combination of purely continuous star formation cannot reproduce the [O~IV]/[Ne~III] ratios observed in most of the star-formation galaxies. We define new regions and demarcation lines in the W10 diagram based on the maximum star-formation contribution derived from these models.

  \item We investigate the optical predictions from our two-zone AGN mid-infrared photoionization grid using the [O~III]/H$\beta$ and [N~II]/H$\alpha$ diagnostic diagram. Given the high density and low ionization of the radiation-bounded component in our two-zone AGN model, we find this region to be a good lower limit to [N~II] emission in agreement with the ionization fraction and the critical density of N~II. In other words, the physical conditions in the [Ne~II] emitting region are not optimal for producing N~II; however, even under these conditions a small fraction of N~II can exist. Therefore we define a new theoretical demarcation line for the minimum AGN contribution to this optical diagnostic diagram. As a result we find that our new classification estimates a higher AGN population by a factor of $\sim$ 1.4 than predictions derived from a semi-empirical division of star-forming galaxies (K03). Moreover, our classification is in good agreement with the semi-empirical starburst demarcation line from \citet{2006MNRAS.371..972S}.

  \item We define a maximum AGN contribution in the [O~III]/H$\beta$ and [N~II]/H$\alpha$ diagram based on a two-zone approximation represented by the combination of any two regions within a range of parameters to maximize the emission of these optical lines. The range of physical parameters relies on our predictions of the ionization fractions and ionic column densities of these ions and is in agreement with typical NLR parameters.
\end{enumerate}

Overall, we use a grid of AGN models to define a new set of demarcation lines that are useful in classifying galaxies. Our theoretical models provide a new perspective into the non-thermal ionizing AGN continuum, and are intended to complement   previous studies that rely solely on the contribution from the stellar ionizing radiation field.


\section*{Acknowledgments}

We thank the referee for very useful comments that improved the paper. This research has made extensive use of the NASA Astrophysics Data System Bibliographic Services and the NASA/IPAC Extragalactic Database (NED), which is operated by the Jet Propulsion Laboratory, California Institute of Technology, under contract with NASA. Funding for SDSS-III has been provided by the Alfred P. Sloan Foundation, the Participating Institutions, the National Science Foundation and the US Department of Energy Office of Science. The SDSS-III web site is http://www.sdss3.org/.

SDSS-III is managed by the Astrophysical Research Consortium for the Participating Institutions of the SDSS-III Collaboration including: the University of Arizona, the Brazilian Participation Group, Brookhaven National Laboratory, Carnegie Mellon University, University of Florida, the French Participation Group, the German Participation Group, Harvard University, the Instituto de Astrof\'isica de Canarias, the Michigan State/Notre Dame/JINA Participation Group, Johns Hopkins University, Lawrence Berkeley National Laboratory, Max Planck Institute for Astrophysics, Max Planck Institute for Extraterrestrial Physics, New Mexico State University, New York University, Ohio State University, Pennsylvania State University, University of Portsmouth, Princeton University, the Spanish Participation Group, University of Tokyo, University of Utah, Vanderbilt University, University of Virginia, University of Washington, and Yale University.

\bibliographystyle{mn2e}
\bibliography{ms}

\begin{thebibliography}{}

\bibitem[\protect\citeauthoryear{{Abel} \& {Satyapal}}{{Abel} \&
  {Satyapal}}{2008}]{2008ApJ...678..686A}
{Abel} N.~P.,  {Satyapal} S.,  2008, \apj, 678, 686

\bibitem[\protect\citeauthoryear{{Ahn}, {Alexandroff}, {Allende Prieto},
  {Anders}, {Anderson}, {Anderton}, {Andrews}, {Aubourg}, {Bailey}, {Bastien}
  \& et al.}{{Ahn} et~al.}{2014}]{2014ApJS..211...17A}
{Ahn} C.~P.,  {Alexandroff} R.,  {Allende Prieto} C.,  {Anders} F.,  {Anderson}
  S.~F.,  {Anderton} T.,  {Andrews} B.~H.,  {Aubourg} {\'E}.,  {Bailey} S.,
  {Bastien} F.~A.,    et al. 2014, \apjs, 211, 17

\bibitem[\protect\citeauthoryear{{Allende Prieto}, {Lambert} \&
  {Asplund}}{{Allende Prieto} et~al.}{2001}]{2001ApJ...556L..63A}
{Allende Prieto} C.,  {Lambert} D.~L.,    {Asplund} M.,  2001, \apjl, 556, L63

\bibitem[\protect\citeauthoryear{{Baldwin}, {Ferland}, {Martin}, {Corbin},
  {Cota}, {Peterson} \& {Slettebak}}{{Baldwin}
  et~al.}{1991}]{1991ApJ...374..580B}
{Baldwin} J.~A.,  {Ferland} G.~J.,  {Martin} P.~G.,  {Corbin} M.~R.,  {Cota}
  S.~A.,  {Peterson} B.~M.,    {Slettebak} A.,  1991, \apj, 374, 580

\bibitem[\protect\citeauthoryear{{Baldwin}, {Phillips} \&
  {Terlevich}}{{Baldwin} et~al.}{1981}]{1981PASP...93....5B}
{Baldwin} J.~A.,  {Phillips} M.~M.,    {Terlevich} R.,  1981, \pasp, 93, 5

\bibitem[\protect\citeauthoryear{{Baskin} \& {Laor}}{{Baskin} \&
  {Laor}}{2005}]{2005MNRAS.358.1043B}
{Baskin} A.,  {Laor} A.,  2005, \mnras, 358, 1043

\bibitem[\protect\citeauthoryear{{Binette}, {Wilson} \&
  {Storchi-Bergmann}}{{Binette} et~al.}{1996}]{1996A&A...312..365B}
{Binette} L.,  {Wilson} A.~S.,    {Storchi-Bergmann} T.,  1996, \aap, 312, 365

\bibitem[\protect\citeauthoryear{{Brinchmann}, {Charlot}, {White}, {Tremonti},
  {Kauffmann}, {Heckman} \& {Brinkmann}}{{Brinchmann}
  et~al.}{2004}]{2004MNRAS.351.1151B}
{Brinchmann} J.,  {Charlot} S.,  {White} S.~D.~M.,  {Tremonti} C.,  {Kauffmann}
  G.,  {Heckman} T.,    {Brinkmann} J.,  2004, \mnras, 351, 1151

\bibitem[\protect\citeauthoryear{{Caputi}}{{Caputi}}{2014}]{2014arXiv1405.7940%
C}
{Caputi} K.~I.,  2014, ArXiv e-prints

\bibitem[\protect\citeauthoryear{{Deo}, {Crenshaw}, {Kraemer}, {Dietrich},
  {Elitzur}, {Teplitz} \& {Turner}}{{Deo} et~al.}{2007}]{2007ApJ...671..124D}
{Deo} R.~P.,  {Crenshaw} D.~M.,  {Kraemer} S.~B.,  {Dietrich} M.,  {Elitzur}
  M.,  {Teplitz} H.,    {Turner} T.~J.,  2007, \apj, 671, 124

\bibitem[\protect\citeauthoryear{{Dicken}, {Tadhunter}, {Morganti}, {Axon},
  {Robinson}, {Magagnoli}, {Kharb}, {Ramos Almeida}, {Mingo}, {Hardcastle},
  {Nesvadba}, {Singh}, {Kouwenhoven}, {Rose}, {Spoon}, {Inskip} \&
  {Holt}}{{Dicken} et~al.}{2014}]{2014arXiv1405.0670D}
{Dicken} D.,  {Tadhunter} C.,  {Morganti} R.,  {Axon} D.,  {Robinson} A.,
  {Magagnoli} M.,  {Kharb} P.,  {Ramos Almeida} C.,  {Mingo} B.,  {Hardcastle}
  M.,  {Nesvadba} N.~P.~H.,  {Singh} V.,  {Kouwenhoven} M.~B.~N.,  {Rose} M.,
  {Spoon} H.,  {Inskip} K.~J.,    {Holt} J.,  2014, ArXiv e-prints

\bibitem[\protect\citeauthoryear{{Farrah}, {Bernard-Salas}, {Spoon}, {Soifer},
  {Armus}, {Brandl}, {Charmandaris}, {Desai}, {Higdon}, {Devost} \&
  {Houck}}{{Farrah} et~al.}{2007}]{2007ApJ...667..149F}
{Farrah} D.,  {Bernard-Salas} J.,  {Spoon} H.~W.~W.,  {Soifer} B.~T.,  {Armus}
  L.,  {Brandl} B.,  {Charmandaris} V.,  {Desai} V.,  {Higdon} S.,  {Devost}
  D.,    {Houck} J.,  2007, \apj, 667, 149

\bibitem[\protect\citeauthoryear{{Ferland}, {Porter}, {van Hoof}, {Williams},
  {Abel}, {Lykins}, {Shaw}, {Henney} \& {Stancil}}{{Ferland}
  et~al.}{2013}]{2013RMxAA..49..137F}
{Ferland} G.~J.,  {Porter} R.~L.,  {van Hoof} P.~A.~M.,  {Williams} R.~J.~R.,
  {Abel} N.~P.,  {Lykins} M.~L.,  {Shaw} G.,  {Henney} W.~J.,    {Stancil}
  P.~C.,  2013, \rmxaa, 49, 137

\bibitem[\protect\citeauthoryear{{Goulding} \& {Alexander}}{{Goulding} \&
  {Alexander}}{2009}]{2009MNRAS.398.1165G}
{Goulding} A.~D.,  {Alexander} D.~M.,  2009, \mnras, 398, 1165

\bibitem[\protect\citeauthoryear{{Groves}, {Dopita} \& {Sutherland}}{{Groves}
  et~al.}{2004a}]{2004ApJS..153....9G}
{Groves} B.~A.,  {Dopita} M.~A.,    {Sutherland} R.~S.,  2004a, \apjs, 153, 9

\bibitem[\protect\citeauthoryear{{Groves}, {Dopita} \& {Sutherland}}{{Groves}
  et~al.}{2004b}]{2004ApJS..153...75G}
{Groves} B.~A.,  {Dopita} M.~A.,    {Sutherland} R.~S.,  2004b, \apjs, 153, 75

\bibitem[\protect\citeauthoryear{{Hao}, {Wu}, {Charmandaris}, {Spoon},
  {Bernard-Salas}, {Devost}, {Lebouteiller} \& {Houck}}{{Hao}
  et~al.}{2009}]{2009ApJ...704.1159H}
{Hao} L.,  {Wu} Y.,  {Charmandaris} V.,  {Spoon} H.~W.~W.,  {Bernard-Salas} J.,
   {Devost} D.,  {Lebouteiller} V.,    {Houck} J.~R.,  2009, \apj, 704, 1159

\bibitem[\protect\citeauthoryear{{Ho} \& {Keto}}{{Ho} \&
  {Keto}}{2007}]{2007ApJ...658..314H}
{Ho} L.~C.,  {Keto} E.,  2007, \apj, 658, 314

\bibitem[\protect\citeauthoryear{{Holweger}}{{Holweger}}{2001}]{2001AIPC..598.%
..23H}
{Holweger} H.,  2001, in {Wimmer-Schweingruber} R.~F.,  ed., Joint SOHO/ACE
  workshop ''Solar and Galactic Composition'' Vol.~598 of American Institute of
  Physics Conference Series, {Photospheric abundances: Problems, updates,
  implications}.
pp 23--30

\bibitem[\protect\citeauthoryear{{Kauffmann}, {Heckman}, {Tremonti},
  {Brinchmann}, {Charlot}, {White}, {Ridgway}, {Brinkmann}, {Fukugita}, {Hall},
  {Ivezi{\'c}}, {Richards} \& {Schneider}}{{Kauffmann}
  et~al.}{2003}]{2003MNRAS.346.1055K}
{Kauffmann} G.,  {Heckman} T.~M.,  {Tremonti} C.,  {Brinchmann} J.,  {Charlot}
  S.,  {White} S.~D.~M.,  {Ridgway} S.~E.,  {Brinkmann} J.,  {Fukugita} M.,
  {Hall} P.~B.,  {Ivezi{\'c}} {\v Z}.,  {Richards} G.~T.,    {Schneider} D.~P.,
   2003, \mnras, 346, 1055

\bibitem[\protect\citeauthoryear{{Kewley}, {Dopita}, {Sutherland}, {Heisler} \&
  {Trevena}}{{Kewley} et~al.}{2001}]{2001ApJ...556..121K}
{Kewley} L.~J.,  {Dopita} M.~A.,  {Sutherland} R.~S.,  {Heisler} C.~A.,
  {Trevena} J.,  2001, \apj, 556, 121

\bibitem[\protect\citeauthoryear{{Kewley}, {Groves}, {Kauffmann} \&
  {Heckman}}{{Kewley} et~al.}{2006}]{2006MNRAS.372..961K}
{Kewley} L.~J.,  {Groves} B.,  {Kauffmann} G.,    {Heckman} T.,  2006, \mnras,
  372, 961

\bibitem[\protect\citeauthoryear{{Kraemer} \& {Crenshaw}}{{Kraemer} \&
  {Crenshaw}}{2000a}]{2000ApJ...532..256K}
{Kraemer} S.~B.,  {Crenshaw} D.~M.,  2000a, \apj, 532, 256

\bibitem[\protect\citeauthoryear{{Kraemer} \& {Crenshaw}}{{Kraemer} \&
  {Crenshaw}}{2000b}]{2000ApJ...544..763K}
{Kraemer} S.~B.,  {Crenshaw} D.~M.,  2000b, \apj, 544, 763

\bibitem[\protect\citeauthoryear{{Kraemer}, {Crenshaw}, {Filippenko} \&
  {Peterson}}{{Kraemer} et~al.}{1998}]{1998ApJ...499..719K}
{Kraemer} S.~B.,  {Crenshaw} D.~M.,  {Filippenko} A.~V.,    {Peterson} B.~M.,
  1998, \apj, 499, 719

\bibitem[\protect\citeauthoryear{{Kraemer}, {Crenshaw}, {Hutchings}, {Gull},
  {Kaiser}, {Nelson} \& {Weistrop}}{{Kraemer}
  et~al.}{2000}]{2000ApJ...531..278K}
{Kraemer} S.~B.,  {Crenshaw} D.~M.,  {Hutchings} J.~B.,  {Gull} T.~R.,
  {Kaiser} M.~E.,  {Nelson} C.~H.,    {Weistrop} D.,  2000, \apj, 531, 278

\bibitem[\protect\citeauthoryear{{Kraemer} \& {Harrington}}{{Kraemer} \&
  {Harrington}}{1986}]{1986ApJ...307..478K}
{Kraemer} S.~B.,  {Harrington} J.~P.,  1986, \apj, 307, 478

\bibitem[\protect\citeauthoryear{{Kraemer}, {Ruiz} \& {Crenshaw}}{{Kraemer}
  et~al.}{1998}]{1998ApJ...508..232K}
{Kraemer} S.~B.,  {Ruiz} J.~R.,    {Crenshaw} D.~M.,  1998, \apj, 508, 232

\bibitem[\protect\citeauthoryear{{Kraemer}, {Trippe}, {Crenshaw},
  {Mel{\'e}ndez}, {Schmitt} \& {Fischer}}{{Kraemer}
  et~al.}{2009}]{2009ApJ...698..106K}
{Kraemer} S.~B.,  {Trippe} M.~L.,  {Crenshaw} D.~M.,  {Mel{\'e}ndez} M.,
  {Schmitt} H.~R.,    {Fischer} T.~C.,  2009, \apj, 698, 106

\bibitem[\protect\citeauthoryear{{Kroupa}}{{Kroupa}}{2001}]{2001MNRAS.322..231%
K}
{Kroupa} P.,  2001, \mnras, 322, 231

\bibitem[\protect\citeauthoryear{{LaMassa}, {Heckman}, {Ptak}, {Schiminovich},
  {O'Dowd} \& {Bertincourt}}{{LaMassa} et~al.}{2012}]{2012ApJ...758....1L}
{LaMassa} S.~M.,  {Heckman} T.~M.,  {Ptak} A.,  {Schiminovich} D.,  {O'Dowd}
  M.,    {Bertincourt} B.,  2012, \apj, 758, 1

\bibitem[\protect\citeauthoryear{{Leitherer}, {Ortiz Ot{\'a}lvaro}, {Bresolin},
  {Kudritzki}, {Lo Faro}, {Pauldrach}, {Pettini} \& {Rix}}{{Leitherer}
  et~al.}{2010}]{2010ApJS..189..309L}
{Leitherer} C.,  {Ortiz Ot{\'a}lvaro} P.~A.,  {Bresolin} F.,  {Kudritzki}
  R.-P.,  {Lo Faro} B.,  {Pauldrach} A.~W.~A.,  {Pettini} M.,    {Rix} S.~A.,
  2010, \apjs, 189, 309

\bibitem[\protect\citeauthoryear{{Leitherer}, {Schaerer}, {Goldader},
  {Gonz{\'a}lez Delgado}, {Robert}, {Kune}, {de Mello}, {Devost} \&
  {Heckman}}{{Leitherer} et~al.}{1999}]{1999ApJS..123....3L}
{Leitherer} C.,  {Schaerer} D.,  {Goldader} J.~D.,  {Gonz{\'a}lez Delgado}
  R.~M.,  {Robert} C.,  {Kune} D.~F.,  {de Mello} D.~F.,  {Devost} D.,
  {Heckman} T.~M.,  1999, \apjs, 123, 3

\bibitem[\protect\citeauthoryear{{Lutz}, {Kunze}, {Spoon} \& {Thornley}}{{Lutz}
  et~al.}{1998}]{1998A&A...333L..75L}
{Lutz} D.,  {Kunze} D.,  {Spoon} H.~W.~W.,    {Thornley} M.~D.,  1998, \aap,
  333, L75

\bibitem[\protect\citeauthoryear{{Mathis}, {Rumpl} \& {Nordsieck}}{{Mathis}
  et~al.}{1977}]{1977ApJ...217..425M}
{Mathis} J.~S.,  {Rumpl} W.,    {Nordsieck} K.~H.,  1977, \apj, 217, 425

\bibitem[\protect\citeauthoryear{{Mel{\'e}ndez}, {Kraemer}, {Armentrout},
  {Deo}, {Crenshaw}, {Schmitt}, {Mushotzky}, {Tueller}, {Markwardt} \&
  {Winter}}{{Mel{\'e}ndez} et~al.}{2008a}]{2008ApJ...682...94M}
{Mel{\'e}ndez} M.,  {Kraemer} S.~B.,  {Armentrout} B.~K.,  {Deo} R.~P.,
  {Crenshaw} D.~M.,  {Schmitt} H.~R.,  {Mushotzky} R.~F.,  {Tueller} J.,
  {Markwardt} C.~B.,    {Winter} L.,  2008a, \apj, 682, 94

\bibitem[\protect\citeauthoryear{{Mel{\'e}ndez}, {Kraemer}, {Schmitt},
  {Crenshaw}, {Deo}, {Mushotzky} \& {Bruhweiler}}{{Mel{\'e}ndez}
  et~al.}{2008b}]{2008ApJ...689...95M}
{Mel{\'e}ndez} M.,  {Kraemer} S.~B.,  {Schmitt} H.~R.,  {Crenshaw} D.~M.,
  {Deo} R.~P.,  {Mushotzky} R.~F.,    {Bruhweiler} F.~C.,  2008b, \apj, 689, 95

\bibitem[\protect\citeauthoryear{{Mel{\'e}ndez}, {Kraemer}, {Weaver} \&
  {Mushotzky}}{{Mel{\'e}ndez} et~al.}{2011}]{2011ApJ...738....6M}
{Mel{\'e}ndez} M.,  {Kraemer} S.~B.,  {Weaver} K.~A.,    {Mushotzky} R.~F.,
  2011, \apj, 738, 6

\bibitem[\protect\citeauthoryear{{M{\"u}ller-S{\'a}nchez},
  {Gonz{\'a}lez-Mart{\'{\i}}n}, {Fern{\'a}ndez-Ontiveros}, {Acosta-Pulido} \&
  {Prieto}}{{M{\"u}ller-S{\'a}nchez} et~al.}{2010}]{2010ApJ...716.1166M}
{M{\"u}ller-S{\'a}nchez} F.,  {Gonz{\'a}lez-Mart{\'{\i}}n} O.,
  {Fern{\'a}ndez-Ontiveros} J.~A.,  {Acosta-Pulido} J.~A.,    {Prieto} M.~A.,
  2010, \apj, 716, 1166

\bibitem[\protect\citeauthoryear{{Netzer} \& {Laor}}{{Netzer} \&
  {Laor}}{1993}]{1993ApJ...404L..51N}
{Netzer} H.,  {Laor} A.,  1993, \apjl, 404, L51

\bibitem[\protect\citeauthoryear{{Osterbrock} \& {Ferland}}{{Osterbrock} \&
  {Ferland}}{2006}]{2006agna.book.....O}
{Osterbrock} D.~E.,  {Ferland} G.~J.,  2006, {Astrophysics of gaseous nebulae
  and active galactic nuclei}

\bibitem[\protect\citeauthoryear{{Osterbrock}, {Tran} \&
  {Veilleux}}{{Osterbrock} et~al.}{1992}]{1992ApJ...389..305O}
{Osterbrock} D.~E.,  {Tran} H.~D.,    {Veilleux} S.,  1992, \apj, 389, 305

\bibitem[\protect\citeauthoryear{{Pereira-Santaella}, {Diamond-Stanic},
  {Alonso-Herrero} \& {Rieke}}{{Pereira-Santaella}
  et~al.}{2010}]{2010ApJ...725.2270P}
{Pereira-Santaella} M.,  {Diamond-Stanic} A.~M.,  {Alonso-Herrero} A.,
  {Rieke} G.~H.,  2010, \apj, 725, 2270

\bibitem[\protect\citeauthoryear{{Rigby}, {Diamond-Stanic} \& {Aniano}}{{Rigby}
  et~al.}{2009}]{2009ApJ...700.1878R}
{Rigby} J.~R.,  {Diamond-Stanic} A.~M.,    {Aniano} G.,  2009, \apj, 700, 1878

\bibitem[\protect\citeauthoryear{{Rigby} \& {Rieke}}{{Rigby} \&
  {Rieke}}{2004}]{2004ApJ...606..237R}
{Rigby} J.~R.,  {Rieke} G.~H.,  2004, \apj, 606, 237

\bibitem[\protect\citeauthoryear{{Rubin}, {Dufour} \& {Walter}}{{Rubin}
  et~al.}{1993}]{1993ApJ...413..242R}
{Rubin} R.~H.,  {Dufour} R.~J.,    {Walter} D.~K.,  1993, \apj, 413, 242

\bibitem[\protect\citeauthoryear{{Rubin}, {Simpson}, {Haas} \&
  {Erickson}}{{Rubin} et~al.}{1991}]{1991ApJ...374..564R}
{Rubin} R.~H.,  {Simpson} J.~P.,  {Haas} M.~R.,    {Erickson} E.~F.,  1991,
  \apj, 374, 564

\bibitem[\protect\citeauthoryear{{Sales}, {Ruschel-Dutra}, {Pastoriza},
  {Riffel} \& {Winge}}{{Sales} et~al.}{2014}]{2014MNRAS.441..630S}
{Sales} D.~A.,  {Ruschel-Dutra} D.,  {Pastoriza} M.~G.,  {Riffel} R.,
  {Winge} C.,  2014, \mnras, 441, 630

\bibitem[\protect\citeauthoryear{{Satyapal}, {Vega}, {Dudik}, {Abel} \&
  {Heckman}}{{Satyapal} et~al.}{2008}]{2008ApJ...677..926S}
{Satyapal} S.,  {Vega} D.,  {Dudik} R.~P.,  {Abel} N.~P.,    {Heckman} T.,
  2008, \apj, 677, 926

\bibitem[\protect\citeauthoryear{{Stasi{\'n}ska}, {Cid Fernandes}, {Mateus},
  {Sodr{\'e}} \& {Asari}}{{Stasi{\'n}ska} et~al.}{2006}]{2006MNRAS.371..972S}
{Stasi{\'n}ska} G.,  {Cid Fernandes} R.,  {Mateus} A.,  {Sodr{\'e}} L.,
  {Asari} N.~V.,  2006, \mnras, 371, 972

\bibitem[\protect\citeauthoryear{{Tommasin}, {Spinoglio}, {Malkan} \&
  {Fazio}}{{Tommasin} et~al.}{2010}]{2010ApJ...709.1257T}
{Tommasin} S.,  {Spinoglio} L.,  {Malkan} M.~A.,    {Fazio} G.,  2010, \apj,
  709, 1257

\bibitem[\protect\citeauthoryear{{Toomre} \& {Toomre}}{{Toomre} \&
  {Toomre}}{1972}]{1972ApJ...178..623T}
{Toomre} A.,  {Toomre} J.,  1972, \apj, 178, 623

\bibitem[\protect\citeauthoryear{{Tremonti}, {Heckman}, {Kauffmann},
  {Brinchmann}, {Charlot}, {White}, {Seibert}, {Peng}, {Schlegel}, {Uomoto},
  {Fukugita} \& {Brinkmann}}{{Tremonti} et~al.}{2004}]{2004ApJ...613..898T}
{Tremonti} C.~A.,  {Heckman} T.~M.,  {Kauffmann} G.,  {Brinchmann} J.,
  {Charlot} S.,  {White} S.~D.~M.,  {Seibert} M.,  {Peng} E.~W.,  {Schlegel}
  D.~J.,  {Uomoto} A.,  {Fukugita} M.,    {Brinkmann} J.,  2004, \apj, 613, 898

\bibitem[\protect\citeauthoryear{{V{\'a}zquez} \& {Leitherer}}{{V{\'a}zquez} \&
  {Leitherer}}{2005}]{2005ApJ...621..695V}
{V{\'a}zquez} G.~A.,  {Leitherer} C.,  2005, \apj, 621, 695

\bibitem[\protect\citeauthoryear{{Veilleux} \& {Osterbrock}}{{Veilleux} \&
  {Osterbrock}}{1987}]{1987ApJS...63..295V}
{Veilleux} S.,  {Osterbrock} D.~E.,  1987, \apjs, 63, 295

\bibitem[\protect\citeauthoryear{{Veilleux}, {Rupke}, {Kim}, {Genzel}, {Sturm},
  {Lutz}, {Contursi}, {Schweitzer}, {Tacconi}, {Netzer}, {Sternberg}, {Mihos},
  {Baker}, {Mazzarella}, {Lord}, {Sanders}, {Stockton}, {Joseph} \&
  {Barnes}}{{Veilleux} et~al.}{2009}]{2009ApJS..182..628V}
{Veilleux} S.,  {Rupke} D.~S.~N.,  {Kim} D.,  {Genzel} R.,  {Sturm} E.,  {Lutz}
  D.,  {Contursi} A.,  {Schweitzer} M.,  {Tacconi} L.~J.,  {Netzer} H.,
  {Sternberg} A.,  {Mihos} J.~C.,  {Baker} A.~J.,  {Mazzarella} J.~M.,  {Lord}
  S.,  {Sanders} D.~B.,  {Stockton} A.,  {Joseph} R.~D.,    {Barnes} J.~E.,
  2009, \apjs, 182, 628

\bibitem[\protect\citeauthoryear{{V{\'e}ron-Cetty} \&
  {V{\'e}ron}}{{V{\'e}ron-Cetty} \& {V{\'e}ron}}{2006}]{2006A&A...455..773V}
{V{\'e}ron-Cetty} M.,  {V{\'e}ron} P.,  2006, \aap, 455, 773

\bibitem[\protect\citeauthoryear{{Weaver}, {Mel{\'e}ndez}, {Mushotzky},
  {Kraemer}, {Engle}, {Malumuth}, {Tueller}, {Markwardt}, {Berghea}, {Dudik},
  {Winter} \& {Armus}}{{Weaver} et~al.}{2010}]{2010ApJ...716.1151W}
{Weaver} K.~A.,  {Mel{\'e}ndez} M.,  {Mushotzky} R.~F.,  {Kraemer} S.,  {Engle}
  K.,  {Malumuth} E.,  {Tueller} J.,  {Markwardt} C.,  {Berghea} C.~T.,
  {Dudik} R.~P.,  {Winter} L.~M.,    {Armus} L.,  2010, \apj, 716, 1151

\end{thebibliography}

\end{document}